\documentclass{article}

\usepackage[final]{neurips_2022_ml4ps}

\usepackage[utf8]{inputenc} 
\usepackage[T1]{fontenc}    
\usepackage{hyperref}       
\usepackage{url}            
\usepackage{booktabs}       
\usepackage{amsfonts}       
\usepackage{nicefrac}       
\usepackage{microtype}      
\usepackage{xcolor}         
\usepackage{comment}
\usepackage{graphicx}
\usepackage{wrapfig}
\usepackage{subfig}
\usepackage{float}
\usepackage{comment}
\usepackage{bm}
\title{Learning Electron Bunch Distribution along a FEL Beamline by Normalising Flows}

\author{%
Anna Willmann \\
Helmholtz Zentrum Dresden-Rossendorf \\
Institute of Radiation Physics\\
\texttt{a.willmann@hzdr.de} \\
\And
    Jurjen Couperus Cabadağ \\
    Institute of Radiation Physics\\
    Helmholtz Zentrum Dresden-Rossendorf \\
\And
    Yen-Yu Chang \\
    Institute of Radiation Physics\\
Helmholtz Zentrum Dresden-Rossendorf \\
\And
    Richard Pausch \\
    Institute of Radiation Physics\\
Helmholtz Zentrum Dresden-Rossendorf \\
\And
    Amin Ghaith \\
    Institute of Radiation Physics\\
Helmholtz Zentrum Dresden-Rossendorf \\
Synchrotron SOLEIL \\
\And
    Alexander Debus \\
    Institute of Radiation Physics\\
Helmholtz Zentrum Dresden-Rossendorf \\
\And
    Arie Irman \\
    Institute of Radiation Physics\\
Helmholtz Zentrum Dresden-Rossendorf \\
\And
    Michael Bussmann\\
    Center for Advanced Systems Understanding\\
    Görlitz\\
\And
    Ulrich Schramm \\
    Institute of Radiation Physics\\
Helmholtz Zentrum Dresden-Rossendorf \\
Technische Universität Dresden\\
\And
    Nico Hoffmann \\
    Institute of Radiation Physics\\
Helmholtz Zentrum Dresden-Rossendorf \\}
    
\begin{document}

\maketitle

\begin{abstract}
    Understanding and control of Laser-driven Free Electron Lasers remain to be difficult problems that require highly intensive experimental and theoretical research. 
    The gap between simulated and experimentally collected data might complicate studies and interpretation of obtained results. 
    In this work we developed a deep learning based surrogate that could help to fill in this gap.
    We introduce a surrogate model based on normalising flows for conditional phase-space representation of electron clouds in a FEL beamline. Achieved results let us discuss further benefits and limitations in exploitability of the models to gain deeper understanding of fundamental processes within a beamline.
\end{abstract}

\section{Introduction and Motivation}
Plasma acceleration processes are comprehensively studied in the recent years as high-energy X-rays find many applications such as medical diagnostics and treatment, chip manufacturing or hardening of material surfaces. X-rays are used in research as well as they can provide atomic scale imaging.

One source of high-brilliance coherent X-rays is free-electron lasers(FELs), where a bunch of relativistic electrons is transported through a sequence of optical elements and passed through an undulator - a magnetic structure with varying polarity, where electrons emmit photons coherently.

Application of conventional particle accelerators to generate relativistic electrons might end up in kilometers in size, what makes them expensive to keep and maintain. 
Laser-plasma accelerators(injectors) are compact and can reduce FEL's costs, however successful application of them requires a precise understanding and control over the process.

\begin{figure}[h!]
  \centering
  \includegraphics[width=0.8\textwidth]{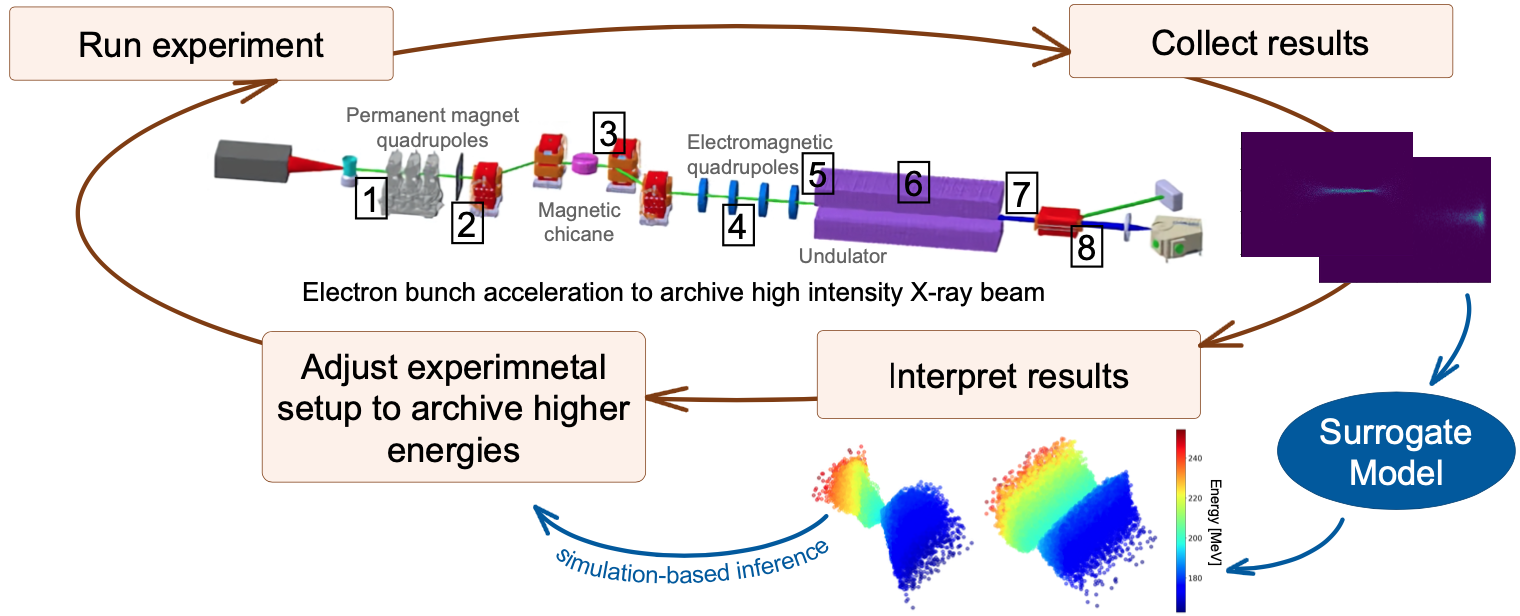}
  \caption{Surrogate model application in laser plasma acceleration research}
  \label{fig:beamline}
\end{figure}

To achieve clear understanding of the process there are performed many experiments supported by theoretical research.
Numerical modeling of a free electron laser consists of multiple parts: simulations of electron acceleration, transport of an electron bunch and radiation in an undulator.
In-situ analysis of a FEL by simulation is limited due to complex workflow and potentially long simulation time.
Detailed description of used equipment of the beamline together with achieved experimental results is given in \citet{hzdr_fel}.

\begin{figure}[h!]
  \centering
  \includegraphics[width=0.8\textwidth]{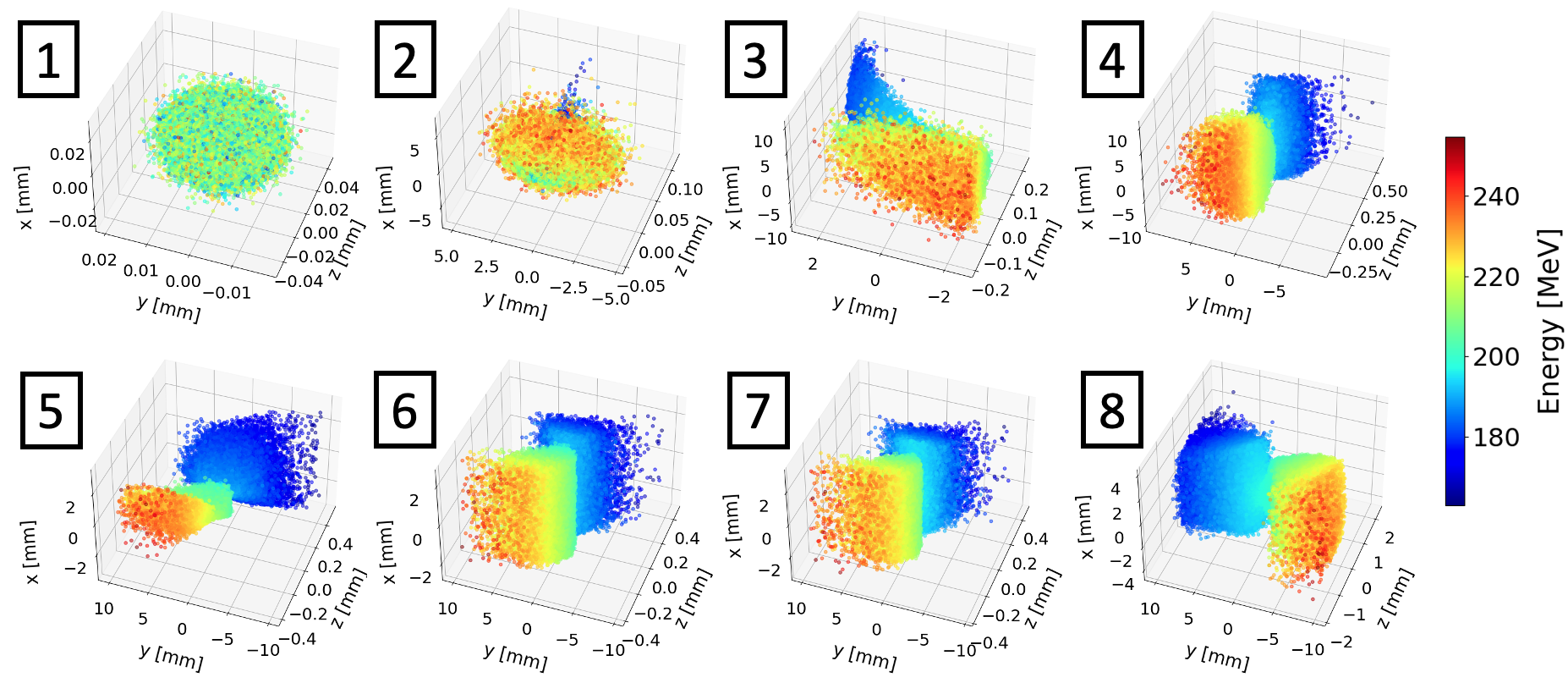}
  \caption{Electron bunch 1 is passed through a sequence of optical elements (fig. \ref{fig:beamline}), each of them is changing the shape of a distribution in order to achieve high coherence of emission in the undulator.}
  \label{fig:tranclouds}
\end{figure}

Figure \ref{fig:beamline} represents a beamline from \citet{coxinel} for a free electron laser with a scatch of the research pipeline.
Electrons are accelerated inside plasma and create a bunch(figure \ref{fig:beamline}, diagnostic screen 1) that is propagated through the manipulation line(figure \ref{fig:beamline}, 2-8) and being transformed as it is shown at figure \ref{fig:tranclouds}.

The operation of the beamline can be monitored by different diagnostics and quantities, among others, spectrometer measurements at positions of diagnostic screens(e.g. diagnostic screen 2 at figure \ref{fig:beamline}) and intensity of radiation. 
We can now tune certain performance criteria of the beamline, e.g. brilliance, stability, higher intensity, by adjusting certain parameters of the injector as well as the facility. 
Alternatively, simulation-based inference(\citet{sbinference}) or data-driven methods for one-step inversion(\citet{Hoffmann2016}) can be used to recover the matching phase-space representation to provide explanation of covered internals of the beamline to advance fundamental research. 
Simulation-based inference, however, relies on a fast surrogate model that can be used for Bayesian inversion by e.g. rejection-sampling (Approximate Bayesian Computation) or Markov Chain Monte Carlo. Advanced techniques in machine learning based surrogate modeling also promises simulation-based design of accelerators which is currently limited due computational challenges for simulation of the governing system(\cite{lehe2020machine}).

Main contribution of this work is therefore a surrogate model that allows for reconstruction of the phase space distribution at different locations of the beamline conditioned on parameters of an accelerated electron bunch by an injector. 

\section{Method}
The non-linear transformation of electrons along a beamline makes it hard to model the corresponding electron distribution determenistically, i.e. by tracking each particle. We therefore propose to learn the distribution of electrons in data-driven fashion by a generative model. There also is an increasing interest in the application of physics-informed neural networks (PINN) for training of surrogate models in physics as demonstrated by (\citet{Stiller2020}). However, we focus on a data-driven surrogate model since PINN increase the computational complexity of the training due to heavy usage of automatic differentiation. There are certain orthogonal state-of-the-art generative models such as GANs(\citet{goodfellow2014generative}), variational autoencoders(\citet{kingma2019introduction}) and normalizing flows(\citet{dinhnice}) while we concentrate on normalizing flows. Compared to variational autoencoders, it has simpler architecture(only one network instead of encoder and decoder) and does not have a loss of information due to compression. 
Negative log-likelihood loss function is easier to balance and optimize compared to the adversarial GANs' loss function.

\textbf{Data.}
We are working on simulated electron bunches propagated through the beamline and observed at 8 diagnostic screens(figure \ref{fig:beamline}), computed by ELEGANT software package(\citet{elegant}).

An electron bunch is represented in phase space, where each particle $\bm{p} \in \mathbb{R}^6$ is described by 6 scalars: 3 spatial coordinates ($x_c,y_c,z_c$), index $c$ is added to indicate a notation as related to the coordinate space, its energy $\gamma$ and per-particle divergence in $X$ and $Y$ dimensions(transverse plane). 
Further we call phase space representation of electron bunch an electron/particle/point cloud, where each electron is characterized by 6 coordinates.
Dimension $Z$(longitudinal coordinates) is a direction of bunch propagation, the bunch is centered around $z_c=0$. 

Each electron bunch consists of approximately 100.000 - 200.000 particles. 
During the transport particles with only a limited range of energies are propagated to the undulator in order to gain coherent emission, for this reason some particles will be filtered out in the beamline. 

In total there are 32 simulations, each simulation consists of 8 electron clouds(1 per diagnostic screen) and identified by statistical emmitance(area occupied by a particle cloud) - $\epsilon_x, \epsilon_y$ and beta function(transverse size of the particle cloud) - $\beta_x, \beta_y$ of the initial electron cloud(figure \ref{fig:beamline}, screen 1, before passing any optical element in the line).

An initial distribution for all simulations defined to be a normal distribution for simplicity of method evaluation.
In future the initial distribution will be extended to more complex shapes.

\textbf{Normalizing Flows.}\label{nfs}
Normalizing Flows(NFs) is a family of generative models, first introduced as a non-parametric method to estimate probability density by \citet{tabak1}. 
Further development of the approach into a deep learning based method was done by L. Dinh, who suggested suitable architectures (\citet{dinhnice}, \citet{dinhrealnvp}) to apply it to higher dimensional data without limitation of network's complexity.

The goal is to find an invertible mapping between an electron $\bm{p}$ and a sampled point $\bm{z} \sim \mathcal{N}(0, 1)$:

$\bm{p}=f^{-1}(\bm{z}, \epsilon_x, \epsilon_y, \beta_x, \beta_y, \bm{s}; \theta)$,

\begin{wrapfigure}[5]{r}{5.5cm}
    \begin{center}
    \vspace*{-12mm}
    \includegraphics[scale=0.2]{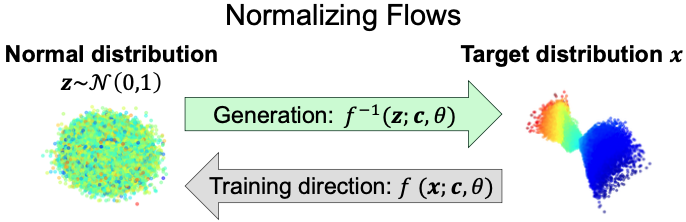}
    \caption{Scheme of normalizing flows}
    \label{fig:nf_appl}
    \end{center}
\end{wrapfigure}

where $\bm{x}=(x_c,y_c,z_c,x_p,y_p,\gamma)$ - phase-space coordinates of each particle 
and $\bm{z}=(z_1, ..., z_6), z_k \sim \mathcal{N}(0, 1), \\ k=1...6$,
$\bm{s}$ - number of a diagnostic screen.
Trainable parameters of the network are defined by $\theta$.
In our work we have a fixed finite number of diagnostic screens and we pass it in the format of a one-hot vector. 
Further $(\epsilon_x, \epsilon_y, \beta_x, \beta_y, \bm{s})$ will be called a condition and denoted by $\bm{c}$. 

Mapping $f$ is trained in the forward direction to minimize the
negative log-likelihood loss function(\citet{cinn}):
$L = max_{\theta}\sum_{i=1}^N -\frac{||f(\bm{p}_i, \bm{c};\theta)||^2_2}{2}+log\left |det\frac{\partial f(\bm{p}_i, \bm{c};\theta)}{\partial \bm{p}_i^T} \right |$

Enhancement of normalizing flows by a masked autoregressive model was suggested by \citet{maf}, where instead of pure samples $\bm{z}$ authors suggest to add dependence on all previous samples:
$\bm{p}_i = \bm{z}_i exp(\mu_i)+\alpha_i$, $\mu_i=f_{\mu_i}(\bm{p}_{1:i-1}, \bm{c};\theta)$, $\alpha_i=f_{\alpha_i}(\bm{p}_{1:i-1}, \bm{c};\theta)$,
where $i$ is a number of a sampled point, 
${f_{\mu_i}, f_{\alpha_i}}$ - MADE networks from \citet{made}.
This architecture allows not to compute $\mu_i$ and $\alpha_i$ recurrently but by a single forward pass for each $i$. $N$ is a total number of samples in a training set.

This re-formulation generalizes the original approach and has a potential to higher expressiveness than conventional NFs.

\begin{wrapfigure}[16]{l}{6.5cm}
    \begin{center}
    \vspace*{-8mm}
    \includegraphics[scale=0.18]{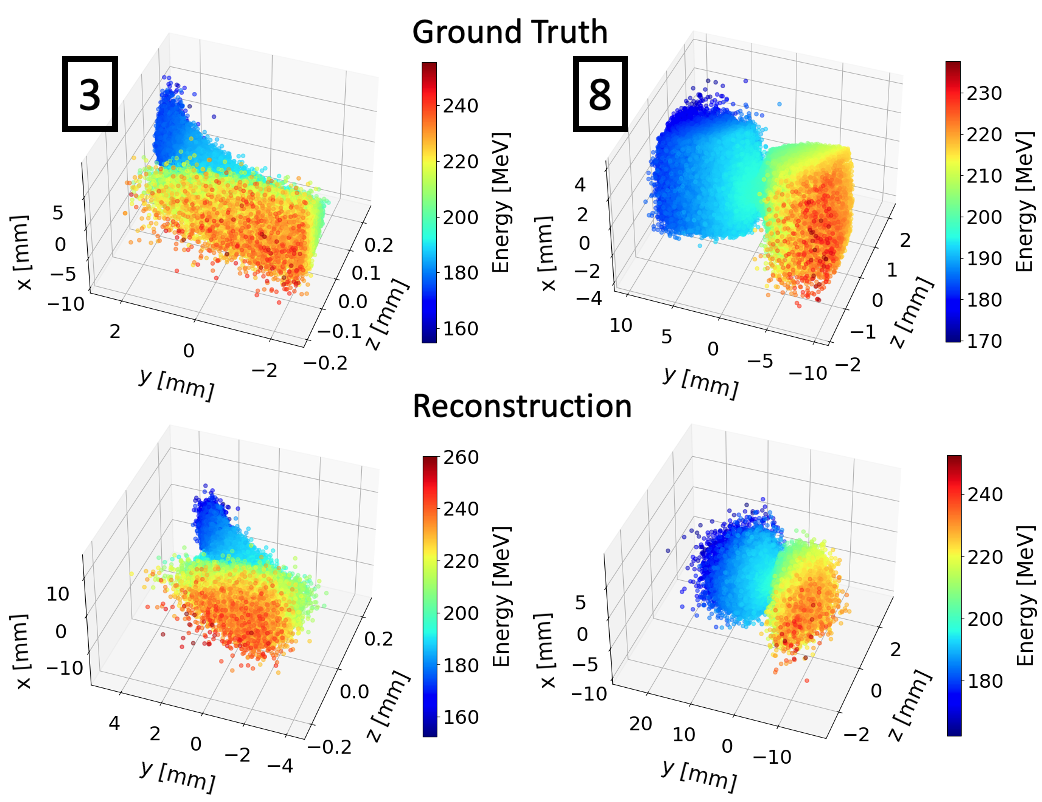}
    \caption{Approximation of point clouds from validation set}
    \label{fig:3d}
    \end{center}
\end{wrapfigure}

\textbf{Model training.} In this work we used two models: a conditional normalizing flow(from \citet{cinn}) and masked autoregressive flow(MAF) from \citet{maf} with a condition. For the conditional NF we use implementation of GLOW blocks(\citet{glow}) provided by FrEIA package(\citet{freia}) and masked autoregressive flow with affine blocks is implemented in \textit{nflows}(\citet{nflows}). Both models consist of 7 coupling blocks with MLP subnetworks.

All passed data are normalized to lie within $[0, 1]$. 
We sample 10.000 particles from each point cloud per iteration, 70 point clouds are processed per batch.
Conditional NF model is trained until convergence over 10.000 iterations using Adam optimizer with learning rate of $10^{-5}$. 
The reached loss on training data is $-14.474$, and on validation data: $-11.705$. The model was trained on a single NVIDIA Tesla P100 16Gb node over 26 hours. 

Masked autoregressive model was trained over 13.750 iterations using Adam optimizer with learning rate of $10^{-3}$. 
The training was distributed(horovod package for distributed learning: \citet{horovod}) over 4 NVIDIA Tesla V100 32Gb nodes. 
Training took 16 hours, reached training loss of $-18.657$ and loss on validation data: $-18.485$. 
Overfitting was not observed.

\begin{wrapfigure}[11]{r}{8.0cm}
    \begin{center}
    \vspace*{-20mm}
    \includegraphics[scale=0.28]{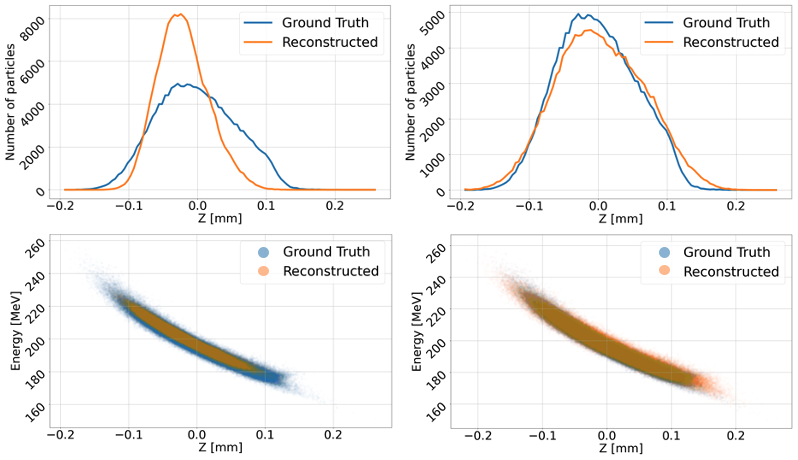}
    \caption{Projections of an approximated point cloud: left column - conditional NFs, right column - MAF}
    \label{fig:slw}
    \end{center}
\end{wrapfigure}

\section{Numerical results} \label{sec:results}

A disadvantage of MAF is a high computational demand, though in the context of our problem input and output have low enough dimensionality.

The models are compared by the following metrics: MMD(\citet{mmd}) and Sinkhorn distance(upper bound of EMD, introduced in \citet{emd}), quantitative results are presented in table \ref{tab:metrics}. Metrics are given in average value(among all clouds in validation set) and the worst reconstructed cloud. 
The masked autoregressive model has shown significantly better reconstruction of point clouds, even the lowest reconstruction quality is higher than a corresponding average value for conditional normalizing flows. 

Figure \ref{fig:3d} demonstrates reconstructed point clouds made by MAF model. 
Shapes of distributions at different positions in the beamline were recongnized and approximated correctly.
Figure \ref{fig:slw} shows projections of point cloud on different planes. 
The top row corresponds to the number of particles per a slice: slices are taken in direction $Z$, each slice has the same width, bottom - 2D projection of particles on plane $Z$ against energy.
As we see, approximated distribution by the conditional NF model does not cover the complete volume of the cloud and most of particles are located around 0, while the MAF model successfully tackles this issue.

\textbf{Limitations.} The dataset has to contain smoothly changing electron clouds to achieve meaningful interpolation.
The number of points in each cloud has to be large enough to represent distribution and allow to sample at each training iteration. 
A low number of particles per cloud cause not stable approximation in a point-wise mapping manner. 
Optimization of hyperparameters plays significant role due to invertibility of the model: 
parameters of subnetworks have to lie within a given range such that results of forward and backward passes do not explode or vanish. 
This limits complexity of transformations, the balance between width of that range and model complexity has to be found. 
An example of this problem one can observe at figure \ref{fig:3d}. 
The reconstruction has certain outliers e.g. in YX-plane: range of values in y axes for ground truth data is around $[-4, 4]$, while in a reconstructed one: $[-20, 20]$. 

\begin{table}
  \caption{Statistical distance on validation data}
  \label{tab:metrics}
  \centering
  \begin{tabular}{lllll}
    \toprule
    Model     & MMD, average & MMD, worst & Sinkhorn, average & Sinkhorn, worst \\
    \midrule
    NF & 0.03293 & 0.06590 & 0.29618 & 0.7245 \\
    MAF & 0.00104 & 0.00180 & 0.08257 & 0.1343 \\
    \bottomrule
  \end{tabular}
\end{table}

\section{Conclusions}
Surrogate modeling can help to optimize experimental research, increase degree of understanding of physical processes in FELs and simplify control over them.
High costs of numerical simulations motivate to use deep learning based methods in order to derive fast and reliable simulation based inference. 
We suggest a masked autoregressive flow based model for conditional generation of electron clouds, that is an important part in the beamline inversion: in order to extract the simulation parameters from diagnostic measurements. 
Results indicate better performance than conditional NFs due to higher expressivity that was theoretically proven by \citet{maf}.  

\section{Impact statement} \label{sec:impact}

The developed surrogate model could significantly accelerate the experimental research of free electron lasers. 
It provides an opportunity to replace simulations by a faster model and optimize parameter space to achieve desired results in experimental companion. 
The negative impact might come from reliability of the model for cases when experimental results do not correspond data used for model training or model was not verified well enough. 
It might cause inconsistent conclusions from collected data during the experiment and bring to wrong interpretations. 
Development of reliable digital twins is a crucial step towards an AI-guided experiments, that can reduce number of experimental runs to explore design space as well as amount of computational resources to produces required simulations. 
It will cause a significant decrease of energy consumption and make research and development of FELs more sustainable.

\newpage

\bibliography{neurips_2022.bib}

\end{document}